\documentclass[aps,preprint,pre]{revtex4}
\usepackage{epsfig}
\usepackage{graphicx}
\usepackage{color}

\def\bea{\begin{eqnarray}}
\def\eea{\end{eqnarray}}
\def\be{\begin{equation}}
\def\ee{\end{equation}}
\def\mez{\hspace{0.5cm}}

\def\be{\begin{equation}}
\def\ee{\end{equation}}
\def\la{\langle}
\def\ra{\rangle}

\usepackage{graphicx}

\begin{document}
\title{ Trapping of cold atoms by the quadrupole force}

\author{Nimrod Moiseyev and  Milan \v{S}indelka }

\affiliation{ Department of Chemistry and Minerva Center of
Nonlinear Physics in Complex Systems, Technion -- Israel Institute
of Technology, Haifa 32000, Israel}

\author {Lorenz S. Cederbaum}

\affiliation{ Theoretische Chemie, Physikalisch-Chemisches
Institut, Universit{\"a}t Heidelberg, D-69120 Heidelberg, Germany}

\begin{abstract}
Cold atoms are traditionally trapped by the dipole force in
periodically spaced potential wells induced by the standing laser
field. We derive here a  theory beyond the conventional dipole
approximation which provides field/atom coupling potential terms
that so far have not been taken into consideration in theoretical
or experimental studies. We show that for some atoms for specific
laser parameters despite the absence of dipole transition  {\it
laser trapping is still possible due to the quadrupole force}.
Illustrative numerical calculations for $Ca$ and $Na$ trapped by the
quadrupole force are given.
%
%
%
\end{abstract}

 \maketitle

\newpage

The possibility of trapping  cold neutral atoms  in periodic
potentials formed by standing laser waves, so called optical
lattices, has opened a new subject of research. For reviews see
for example
Refs.~\cite{CT-phillips,CT-optical-lattices,Jenssen-optical-lattices,phillips-optical-lattices,
Raizen-optical-lattices} and references therein.  Laser cooling
and the formation of optical lattices were explained by the
ability to influence the translational motion of neutral atoms by
the laser induced dipole force~\cite{phillips-optical-lattices}.
Until now, the formation of optical lattices has been studied
theoretically and experimentally exclusively  due to the dipole
interactions. An important question arises, however, on what
happens when the dipole transition matrix element vanishes due to
a symmetry property of the studied atomic system. This is the
case, for example, when the laser frequency is detuned from an
excited state which is associated with a $d$-type symmetry orbital
whereas in the ground state the valence shell electron occupies an
$s$-type symmetry orbital. Does this imply that in such a case
atoms can not be trapped because one can not form an optical
lattice ?

The purpose of the present study is to demonstrate that it is
possible to trap atoms even when the dipole transition between the
two relevant electronic levels coupled by the laser is prohibited
by symmetry. We derive here the Hamiltonian of optical lattices
without imposing the dipole approximation, and study the
possibility of generating optical lattices due to the quadrupole
contributions to the trapping potential which until now have been
neglected. Two illustrative examples are presented explicitly.
Namely, we show that by using realistic laser parameters the
quadrupole force induced optical lattice can be formed and trap
$Ca$ and $Na$ atoms.

Let us consider an atom interacting with a linearly polarized
laser light. The light propagation direction is chosen to be $x$,
and the field is assumed to oscillate along the $z$-direction. For
the sake of simplicity and without loss of generality we describe
the field free atomic Hamiltonian by an effective one electron
model Hamiltonian, ${\bf H}_{\rm el}^{FF}$. The formulation given
here can be extended in a straightforward manner to the general
case of a many electron atom (or ion) interacting with an
arbitrary (not necessarily linearly polarized) light pulse. Such a
generalization will be elaborated elsewhere~\cite{in-preparation}.
By switching from the laboratory frame coordinates to the relative
$ \vec{q}=(x,y,z)$ and center-of-mass (c.m.) $\vec{R}=(X,Y,Z)$
coordinates, one obtains the full Hamiltonian of an effective one
electron atom in a linearly polarized laser field,
\begin{equation}
\label{HAM}
 {\bf H}(t) =  -\frac{\hbar^2}{2M}\nabla_{{\vec R}}^2 +
 {\bf H}_{\rm el}^{FF}(\vec{q}) \; + \; {\bf V}_{coup}(\vec{q},{\vec
 R},t)
\end{equation}
with an interaction term $ {\bf V}_{coup}(\vec{q},{\vec R},t)=
{e}({m_ec})^{-1}A_{m_e}{\hat p}_z  +
{e^2}(2m_ec^2)^{-1}A^{(2)}_{m_e}  +  {e}(cM)^{-1}A_M{\hat{P}}_Z  +
{e^2}(2(M-m_e)c^2)^{-1}A^{(2)}_{M}$. Here, the $A$-factors are
defined in terms of the time, relative coordinate, and
c.m.~coordinate dependent vector potential $A(x,X,t)$, as follows:
$A_{m_e} = [A(x\xi_1+X,t) + (\xi_2/\xi_1)A(-x\xi_2+X,t)] \; ; \;
A_M = [A(x\xi_1+X,t)-A(-x\xi_2+X,t)]\; ; \; A^{(2)}_{m_e}   =
A^2(x\xi_1+X,t)$ and $A^{(2)}_{M} = A^2(-x\xi_2+X,t)$. The symbols
$m_e$ and $M$ stand, respectively, for the mass of the electron
and the atom, and parameters $\xi_1=(M-m_e)/M \, , \,
\xi_2=m_e/M\,$. When the field intensity $I$ is weak ($I$ is
proportional to the squared field amplitude) the linear
interaction terms $A_{m_e}(x,X,t){\hat p}_z$ and
$A_{M}(x,X,t)\hat{P}_Z$ become dominant over the quadratic
interaction terms $A_{m_e}^{(2)}$ and $A_{M}^{(2)}$ so that the
latter two contributions can be neglected. Moreover, since $M \gg
m_e$ it is clear that $\xi_1 \approx 1$ and thus the dominant term
in ${\bf V}_{coup}(\vec{q},{\vec R},t)$
is the first one. In summary, we conclude
that the coupling between the c.m.~and relative coordinates can be
considered as \be \label{Vcoup_simple}
   {\bf V}_{coup}(\vec{q},{\vec R},t) \; = \;
   {e}(m_ec)^{-1}A_{m_e}(x,X,t){\hat p}_z \mez .
\ee

In standard treatments of the atom-field Hamiltonian the
dependence of the vector potential on the relative coordinate has
been neglected \cite{HAMILTONIAN}. That is, $A(x+X,t)\sim A(X,t)$.
Here, we do {\it not} use this so called dipole approximation.
Instead, the spatial dependence of the vector potential is taken
into account, via the Taylor expansion around $x=0$.
An important question arises on what should be taken as the small
parameter in such a Taylor series expansion.
In the case when the laser intensity is low, the associated time
dependent electronic wavefunctions are only slightly different
from their field-free counterparts. Therefore, they possess
exponentially small values for $q>a_0$, where  $a_0$ stands for
the size of the atom. For weak fields, as in our case, $a_0$ is
essentially equal to the radius of the field free atom, $r_0$. Of
course,  in the presence of very strong fields $a_0$ varies with
the intensity as the electron oscillates with the field. Roughly
speaking we may think that the relevant radius of the atom is then
$r_0+ \alpha_0$ where $\alpha_0 = (eA_0)^2/(c^2m_e\omega_L)$ and
$A_0$ is the field amplitude \cite{KRAMMER-HENNEVERGER}. In order
to find a qualitative criterion for convergence of our Taylor
series expansion of $A(x+X,t)$, the above estimated atomic radius
$a_0$ should be compared with the wavelength $\lambda$ of the
laser. Clearly, $\lambda$ should get sufficiently large values
such that $\lambda \gg a_0\,$. The small parameter of the
considered Taylor series expansion is thus found to be $k_La_0$,
where $k_L=2\pi/\lambda\,$.

Following the above discussion, when the wavelengths which
constitute the dominant part of the laser pulse are sufficiently
large, $\lambda \gg a_0$, the two leading order terms in the
Taylor series expansion of the vector potential are given by
\begin{eqnarray}
 {A(x+X,t)}\sim
 A(X,t)+x \, {\partial A(X,t)}/{\partial X}.
\label{expanded-A}
\end{eqnarray}
The first term in Eq.~(\ref{expanded-A})
 is the dipole term,
whereas the second one is the quadrupole term which is taken here
into consideration for the first time in contrast to other
theoretical studies published up to now. Substituting
Eq.~(\ref{expanded-A}) into formula (\ref{Vcoup_simple}) one finds
that
\begin{equation}
{\bf V}_{coup}(\vec{q},{\vec R},t) = {\bf V}_{\rm dip}(z,X,t) +
{\bf V}_{\rm qd}(z,x,X,t) \label{COUPL-POT}
\end{equation}
where the dipole and quadrupole potential terms that couple the
internal (electronic) and c.m.~degrees of freedom are respectively
given by, ${\bf V}_{dip}(z,X,t)=(e/c)m_e^{-1}A(X,t){\hat p}_z$ and
${\bf V}_{qd}(z,x,X,t)=(e/c)m_e^{-1}[\partial{A(X,t)}/\partial X]
x{\hat p}_z$.

Since the field oscillates periodically in time the Hamiltonian is
time periodic with the period $T=2\pi/\omega_L$ where $\omega_L$
is the laser frequency. In such a case the solutions of the time
dependent Schr\"odinger equation are the Floquet solutions,
similarly as the Bloch states are the solutions of the
Schr\"odinger equation for spatially periodic potentials. Within
the framework of the Floquet theory, the time averaged optical
potential, ${\bar V}_{opt}(X)$, is associated with the
quasi-energy eigenvalue of the Floquet operator, ${\bf {\cal
H}}_F=-i\hbar{\partial}/{\partial t} + {\bf H}_{\rm
el}^{FF}(\vec{q}) + {\bf V}_{coup}(\vec{q},{\vec R},t)$, (for
Floquet formalism see for example Ref.\cite{FAISAL}),
\begin{equation}
{\bf {\cal  H}}_F \Phi_{opt}^{QE}(\vec {q},X,t) = {\bar
V}_{opt}(X) \Phi_{opt}^{QE}(\vec {q},X,t)\,.
\label{OPTICAL-POTENTIAL}
\end{equation}
The eigenvalue solution of Eq.~(\ref{OPTICAL-POTENTIAL}) provides
the definition of the optical lattice potential. Strictly
speaking, optical potentials for the atoms exist only in the
adiabatic approximation where one can separate the atomic center
of mass motion from the electronic motion (similarly to the
Born-Oppenheimer or adiabatic approximation which separates the
nuclear and electronic motions in molecules). Within the adiabatic
approximation, our claim that the optical potential equals to an
eigen-energy of the Floquet operator is {\it exact}. Of course,
there are infinitely many eigen-energies and hence optical
potentials (similarly to molecules where in each electronic state
there is a potential for the vibrations). In equation
(\ref{OPTICAL-POTENTIAL}) we have chosen that eigen-energy which
correlates to the field-free atomic ground state. In the weak
field limit we will show below that in the dipole case the same
expression as that used in the literature results, whereas in the
quadrupole case  a novel expression for the quadrupole force
induced optical lattice potential is obtained. Note however that
for sufficiently strong fields, the literature expression (which
is found to be a second order perturbation theory result) fails
whereas the eigen-energy of the Floquet operator still provides an
exact optical potential.

Let us derive now an analytical expression for the optical lattice
potential which includes the leading order term beyond the dipole
approximation. From Eq.~(\ref{expanded-A})
 it is clear that the only information we
need in order to go beyond the dipole approximation is  $A(X,t)$
and its first order derivative with respect to the c.m.~coordinate
$X$. For atoms in a standing laser beam  $ A(X,t) = A_0 \,
\cos(k_LX)\cos(\omega_Lt) \label{standing-waves}$, where
$\omega_L$ is the corresponding laser frequency, and $k_L =
\omega_L/c\,$. Suppose that the atomic ground state $|{\rm g}\ra$
(with energy ${\cal E}_g^0$) is significantly coupled by the laser
light only to a single excited electronic state $|{\rm e}\ra$
(with energy ${\cal E}_e^0 = {\cal E}_g^0 + \hbar\omega_{\rm
atom}$). By following  the "traditional" approach based upon the
adiabatic elimination scheme~\cite{HAMILTONIAN}, we obtain the
time dependent optical lattice potential (including the new term
beyond the dipole approximation), ${\bf V}_{opt}(X,t)  = |{\la
{\rm e} | {\bf V}_{\rm dip}(z,X,t)+{\bf V}_{\rm qd}(z,x,X,t) |
{\rm g} \ra}|^2/ ({\hbar\Delta_L})$, where, $\Delta_L =
\omega_{\rm atom} - \omega_L$. The characteristic timescale of the
center of mass dynamics is by several orders of magnitude smaller
than the time period of an optical cycle, $T =
\lambda/c=2\pi/\omega_L$. Therefore, the time averaged optical
lattice potential reads as $ \bar{V}_{opt}(X) = ({1}/{T} )\,
\int_{0}^{T} V_{opt}(X,t) \; dt $. In case that the laser detuning
with respect to the atomic frequency is such that it couples an
$s$-type atomic state $|{\rm g}\ra$ with a $p$-type atomic state
$|{\rm e}\ra$, the quadrupole contribution to $\bar{V}_{opt}(X)$
vanishes, and the optical lattice potential is reduced to the
expression \be \label{V_dip_final}
   \bar{V}_{opt}^{dip}(X) = {\hbar|\Omega_{\rm dip}|^2}({4\Delta_L})^{-1} \cos^2(k_L X)
\ee where the Rabi frequency is given by $ \Omega_{\rm dip} = ({e
\, A_0})/({\hbar c \, m_e}) \, \la {\rm e} | {\hat p}_z| {\rm g}
\ra. $ This is, of course, the well known conventional expression
for the optical lattice potential formed by the dipole force
\cite{HAMILTONIAN}.

Let us consider now the case when the laser detuning is off
resonance but not from a $p$-type atomic state but from a $d$-type
atomic state. Consequently,  the force exerted on the atom is of
quadrupole nature since the dipole contribution vanishes. Note
that for Calcium, for example, the {\it lowest} excitation is from
an $s$ to a $d$ orbital. Introducing the ''generalized'' Rabi
frequency $\Omega_{qd} = ({e \, A_0})/({\hbar c \, m_e}) \; k_L \;
\la {\rm e} | x {\hat p}_z | {\rm g} \ra$ we obtain the formula
\begin{equation}
\label{V_qd_final}
   {\bar V}_{opt}^{qd}(X)  = {\hbar\left|\Omega_{\rm
   qd}^{}\right|^2}({4\Delta_L})^{-1}
   \, \sin^2(k_L X)
\label{dip}
\end{equation}
which exhibits remarkable similarity to the standard expression
(\ref{V_dip_final}) for the dipole optical lattice potential.
Before proceeding further, let us recall once again that both
relations (\ref{V_dip_final}) and (\ref{V_qd_final}) can be
derived directly by applying the second order perturbation theory
to the Floquet eigenproblem
(\ref{OPTICAL-POTENTIAL})\cite{in-preparation}.

Our theoretical analysis concludes with a very brief discussion of
the as yet neglected incoherent effects (friction force,
spontaneous emission) which are caused by the quantized nature of
the electromagnetic field. It is known \cite{Drake} that the
mentioned dissipation processes can be safely ignored when
$\Delta_L \gg \Gamma$, where $\Gamma$ is the natural linewidth of
the excited electronic state. For the dipole $|{\rm e}\ra \to
|{\rm g}\ra$ transitions,
\be \label{Gamma_dip}
   \Gamma_{\rm dip}  =
   {4e^2\omega_{\rm atom}}({3m_e^2\hbar c^3})^{-1}
|\la {\rm e} | {\hat p}_z \, | {\rm g} \ra|^2\,,
   \mez
\ee
see, e.g.,~Ref.~\cite{spontaneous-emission}. The derivation
leading to Eq.~(\ref{Gamma_dip})  can be extended in a
straightforward manner to cover also the situation when the
transition $|{\rm e}\ra \to |{\rm g}\ra$ occurs solely via the
quadrupole coupling. The final result is then given by formula
 \be
\label{Gamma_qd}
   \Gamma_{\rm qd}  =  {4
   e^2 \omega_{\rm atom}^3}({5m_e^2 \hbar c^5})^{-1}
   |\la {\rm e} | x {\hat p}_z | {\rm g} \ra |^2 \mez .
\ee

Let us apply now our above theoretical results and study the
feasibility of trapping $Na$ and $Ca$ atoms by the quadrupole
force in the periodically spaced potential wells induced by the
standing laser waves. Our model Hamiltonian consists of the ground
S state, $|E_S\ra$, and the lowest excited P and D states,
$|E_P\ra$ and $|E_D\ra$. The corresponding energy levels for $Ca$
and $Na$ were taken from the NIST atomic database.
The field free electronic Hamiltonian is a diagonal $3 \times 3$
matrix, ${\bf E}_e$, where the diagonal terms are the mentioned
atomic energies $E_S,E_P,E_D$. Following
Eq.~(\ref{OPTICAL-POTENTIAL}), the interaction potential induced
by the linearly polarized standing laser field is described by the
symmetric $3 \times 3$ matrix ${\bf V}_{coup}(X,t)$, where the
diagonal matrix elements vanish. The off diagonal matrix elements
are connected either with the dipole or the quadrupole
transitions, and take an explicit form:
$(V_{coup})_{1,2}=eA_0/(m_e c) \la E_S|{\hat
p}_z|E_P\ra\cos(k_LX)\cos(\omega_Lt)$, $(V_{coup})_{2,3}=eA_0/(m_e
c) \la E_P|{\hat p}_z|E_D \ra \cos(k_LX)\cos(\omega_Lt)$ and
$(V_{coup})_{1,3}=-eA_0/(m_e c)k_L \la E_s|{x\hat p}_z|E_D \ra
\sin(k_LX) \cos(\omega_Lt)$.
The corresponding dipole and quadrupole transition matrix elements
for $Na$ and $Ca$ have been calculated using equations
(\ref{Gamma_dip}) and (\ref{Gamma_qd}) from available
theoretical/experimental data for the associated transition rates.
More specifically, the dipole transition linewidths $\Gamma_{\rm
dip}^{Na}$ and $\Gamma_{\rm dip}^{Ca}$ are extracted from the NIST
database,
 while the quadrupole terms
$\Gamma_{\rm qd}^{Na}$ and $\Gamma_{\rm qd}^{Ca}$ are taken from
the literature \cite{Na-quadrupole,Ca-quadrupole}. Referring to
previous discussion of Eq.~(\ref{OPTICAL-POTENTIAL}), we recall
that there is a single quasi-energy eigenvalue of the Floquet
hamiltonian ${\cal H}_F = {\bf E}_e + {\bf V}_{coup}(X,t) -
i\hbar(\partial/\partial t)$ which is associated with the field
free ground state and which is interpreted physically as the
appropriate numerically exact optical lattice potential.

In the first step of our calculations we computed the variation of
the depth of the optical lattice potential well  $V_{\rm max} =
{\rm Max}[{\bar V}_{opt}(X)]$ for $Na$ atoms as a function of the
laser intensity, $I\; [{\rm W}/{\rm cm}^2] \propto ( k_L A_0/2 )^2
$ for the laser parameters as were used recently in the experiment
of Phillips group~\cite{phillips}. That is, $Na$ atoms are trapped
in a one dimensional optical lattice induced by a laser with
$\lambda=589 \,nm$, and the detuning  from the first excited P
state of $Na$ is equal to $\Delta_L=14$ GHz. Our calculations show
that for $I < 100$ W/cm$^2$ the analytical results derived from
the second order perturbation theory, Eq.~(\ref{V_dip_final}), are
in a remarkable agreement with the numerically exact values.
Therefore, indeed within this region of field intensities the
optical lattice is formed exclusively by the dipole force.

The situation is entirely  different when the laser wavelength is
taken to be $\lambda=343\,nm$ and the detuning from the first
excited $D$ state is $\Delta_L=1$ GHz. From Fig.~1 one can see
that the quadrupole force induced optical lattice potential
predicted by Eq.~(\ref{V_qd_final}) is now dominant and in
complete agreement with the numerically exact solutions of
Eq.\ref{OPTICAL-POTENTIAL}, for the entire range of displayed
field intensities. The dipole force contribution to the formation
of $Na$ optical lattice as calculated from Eq.~(\ref{V_dip_final})
is, for the considered laser parameters, by one order of magnitude
smaller than the contribution of the quadrupole force. To achieve
for the quadrupole induced optical lattice the same depth as being
produced for the dipole induced optical lattice potential in the
NIST experiment~\cite{phillips} one must increase the field
intensity by 4 orders of magnitude. This is not a practical
intensity  in nowadays technology. However, by increasing the
intensity by less than 2 orders of magnitude, the depth of the
quadrupole induced optical lattice potential is smaller just by
about 2 orders of magnitude than in the NIST experiments. Such a
potential can trap $Na$ atoms provided that they are cooled down
to less than 0.1$\mu$K. The required temperature is sufficiently
low to keep the atoms trapped in the quadrupole induced optical
lattice potential well since $k_BT< V_{\rm max}$.

Let us now turn to the case of $Ca$ atoms where the first excited
state is a $D$ state. The results of our calculations which are
presented in Fig.~2 show that when the laser light is detuned
slightly from the S $\to$ D transition ($\lambda = 457$ nm,
$\Delta_L = 70$ kHz) it can trap the  $Ca$ atoms solely by
quadrupole interactions. The $Ca$ atoms must, however, be
pre-cooled. It is known that already today ultracold calcium can be
prepared at 10 $\mu K$ ~\cite{Ca_ultracold}. At this temperature the
$Ca$ atoms can be trapped by the quadrupole lattice potential using
a laser intensity of 10 W/cm$^2$. Other values of interest: In order
to form a quadrupole induced optical lattice potential for field
intensities which are smaller than $100$ W/cm$^2$, the $Ca$ atoms
should be cooled to a temperature which is smaller than about $30\mu
K$. For the intensity of $I=1$ W/cm$^2$ the $Ca$ atoms will be
trapped by the quadrupole force when their temperature is smaller
than $3\mu K$. In passing we note that the used value of detuning
$\Delta_L = 70$ kHz is relatively small compared to that of Fig.~1,
but still remains more than 200 times larger than the natural
linewidth $\Gamma \approx 300$ Hz of the upper D state of $Ca$
\cite{Ca-quadrupole}. Therefore, the spontaneous emission effects
can be safely neglected here. This statement has been confirmed also
by an additional numerical calculation \cite{in-preparation}.

Above illustrative examples of course do not restrict the range of
possible applications just to alkali metal or alkali earth atoms.
Similar kind of quadrupole trapping can be exploited e.g.~in the
case of rare gases, which have been laser cooled and Bose condensed
as well (see for example Ref.~\cite{rare_gases}).

In conclusion, we would like to re-stress that our formulation based
upon the Floquet eigenstates is exact for all field intensities.
This method provides a playground for a rational design of optical
lattice potentials induced by the dipole or the quadrupole forces or
by any combination of them. Design of quadrupole optical lattices
might prove to be important in every practical situation where a
specific lattice constant is required and where no dipole transition
possesses a suitable wavelength. Such applications may include
e.g.~interference experiments with wavepackets of cold atoms, or an
enhancement of the high harmonic generation efficiency via
constructive interference/phase matching. Besides trapping of cold
atoms in optical lattices, the theoretical framework described in
the present paper can be used also for other purposes, e.g.~for
studying an effect of quadrupole interaction on the performance of
atomic clocks (cf.~Ref.~\cite{clocks}).

\begin{acknowledgments} This work was supported in part by the by the Israel Science
Foundation (grant no. 1152/04) and by the Fund of promotion of
research at the Technion. One of us (NM) wish to thank Avner
Fleischer for his most helpful comments and suggestions. Support
by the DIP is gratefully acknowledged.
\end{acknowledgments}

\newpage
\begin{figure} \begin{centering}
\includegraphics[angle=-90]{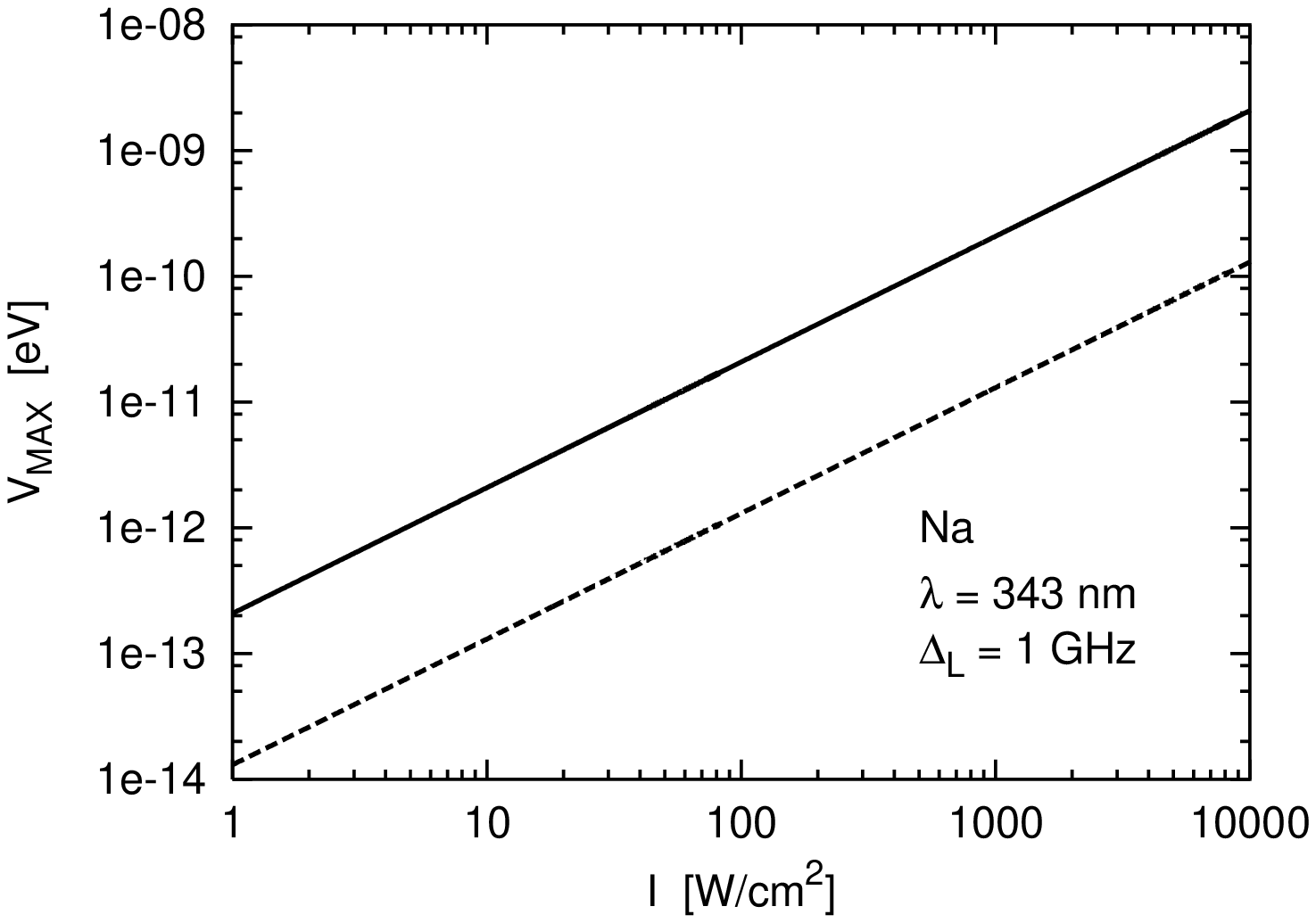} \caption{
The depth $V_{\rm max}$ of the optical lattice potential well for
$Na$ atoms plotted as a function of the laser intensity $I$. The
laser parameters, $\lambda=343$ nm and $\Delta_L = 1$ GHz,
correspond to the transition from the S ground state to the lowest
D state of $Na$. The solid line represents the numerically exact
solutions of Eq.~(\ref{OPTICAL-POTENTIAL}) which coalesces to all
significant digits with the second order perturbation theory
contribution of the quadrupole forces, see Eq.~(\ref{V_qd_final}).
The dashed line shows the contribution of dipole forces
calculated from Eq.~(\ref{V_dip_final}). } \label{Fig1}
\end{centering}
\end{figure}

\newpage \begin{figure} \begin{centering}
\includegraphics[angle=-90]{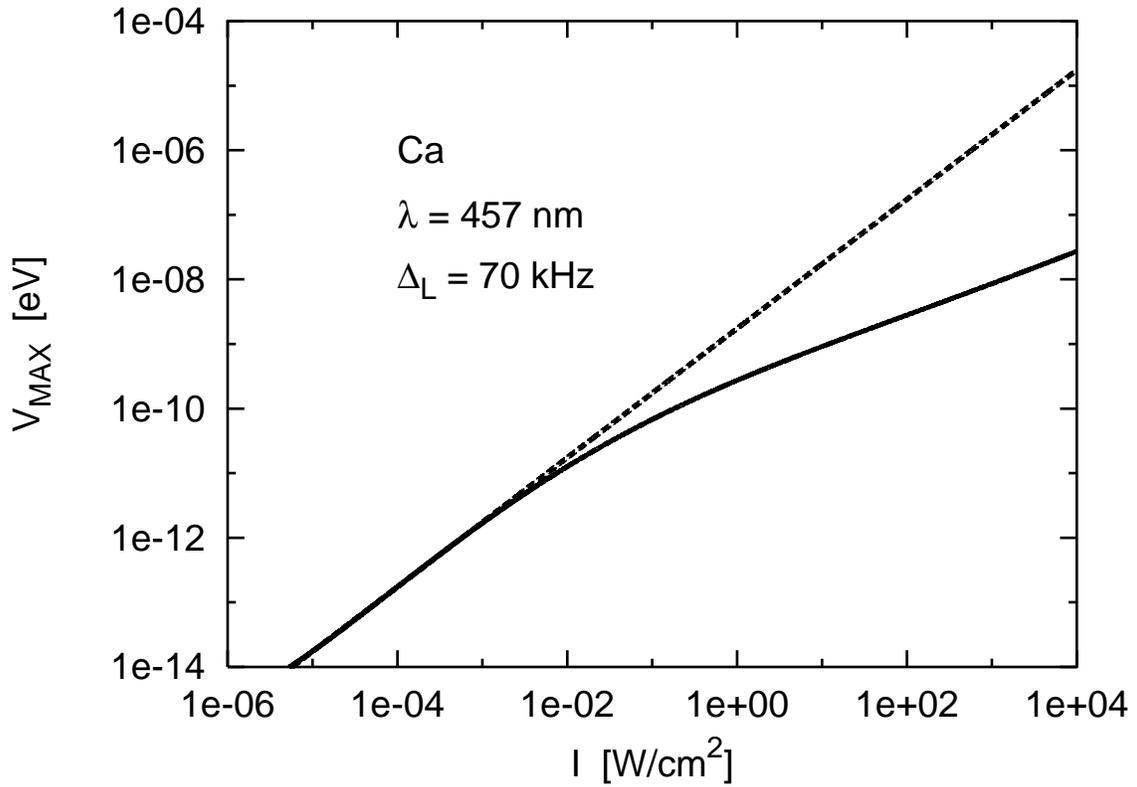} \caption{
The depth $V_{\rm max}$ of the optical lattice potential well for
$Ca$ atoms plotted as a function of the laser intensity $I$. The
laser parameters, $\lambda=457$ nm and $\Delta_L = 70$ kHz,
correspond to the transition from the S ground state to the first
excited D state of $Ca$. The solid line stands here for the
numerically exact optical lattice potential as obtained from the
numerical solution of Eq.~(\ref{OPTICAL-POTENTIAL}). The dashed
line shows the contribution of quadrupole forces accounted within
the second order perturbation theory, see Eq.~(\ref{V_qd_final}).}
\label{Fig2}
\end{centering}
\end{figure}

\end{document}